# Thermodynamic Properties of Gaseous Plasmas in Zero-Temperature Limit

Igor Iosilevskiy

*Moscow Institute of Physics and Technology (State University)*

ilios@orc.ru

Limiting structure of thermodynamic functions of gaseous plasmas is under consideration in the limit of extremely low temperature and density. Remarkable tendency, which was claimed previously [1][2][3], is carried to extreme. The point is that the discussed limit ($T \to 0$; $n \to 0$) is carried out at fixed value for chemical potential of electrons ($\mu_{el}$ = *const*) or "atoms" ($\mu_a = Z\mu_e + \mu_i$ = *const*) or "molecule" ($\mu_{m \leftrightarrow 2a} = 2\mu_a$ = *const*) *etc*. In this limit both equations of state (EOS) thermal and caloric ones, obtain almost identical *stepped* structure ("ionization stairs" [3]) when one uses special forms for exposition of these EOS as a function of electron chemical potential: i.e. $PV/RT$ for thermal EOS and $U - (3/2)PV$ for caloric EOS *vs.* $\mu_{el}$. Examples of this limiting structure are exposed at figures 1 and 2 for thermal and caloric EOS of lithium and helium plasmas [4][6]. For rigorous theoretical proof of existing the limit, which is under discussion (Saha-limit) in the case of hydrogen see [7][8] and references therein.

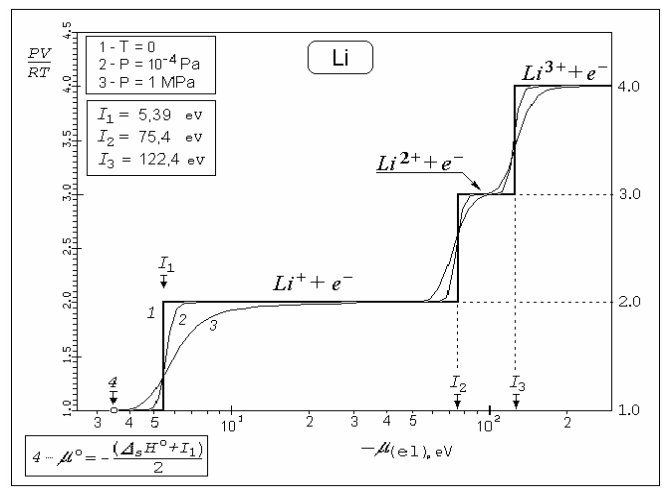

Figure 1. Thermal EOS of lithium plasma in quasi-chemical limit (figure from [6][4]). Compressibility factor $PV/RT \equiv P/(n_{Li}kT)$ as a function of (negative) value of electron chemical potential. <u>Notations</u>: *1* – isotherm $T = 0$; *2, 3* – isobars ($P$ = *const*); *arrows* – elements of "*intrinsic energy scale*" for lithium: $I_1$, $I_2$, $I_3$ – lithium ionization potentials; *4* – saturation vapour boundary $\{(\mu_e)^0 = -(\Delta_s H^0 + I_1)/2\}$. (Isobars *2, 3* are calculated via code SAHA-IV [9] with neglecting of equilibrium radiation contribution)

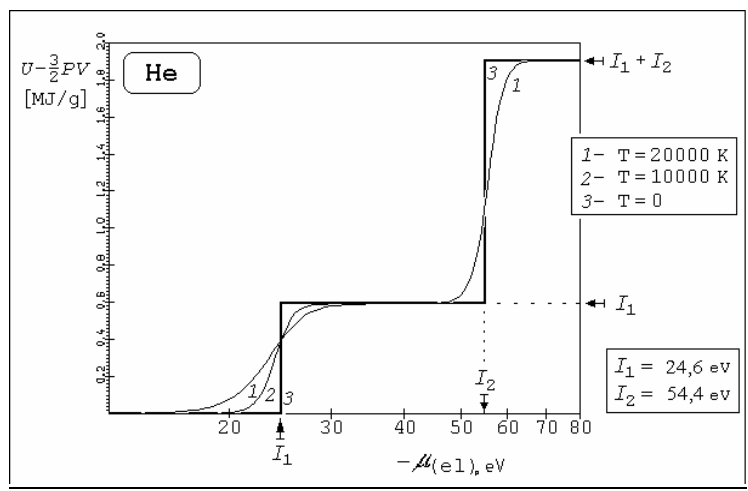

Figure 2. Caloric EOS of helium plasma in quasi-chemical limit (figure from [6] [10]). Complex $U - (3/2)PV$ as a function of (negative) value of electron chemical potential. <u>Notations</u>: *1* – isotherm $T$ = 20 000 K; *2* – $T$ = 10 000 K; *3* – $T$ = 0 K; *arrows* – elements of helium "*intrinsic energy scale*": $I_1$, $I_2$ – helium ionization potentials; (Isotherms *1, 2* are calculated via code SAHA-IV [9] with neglecting of equilibrium radiation contribution)

The same *stepped* structure appears in the zero-temperature limit in any molecular gases, for example hydrogen [4][6].

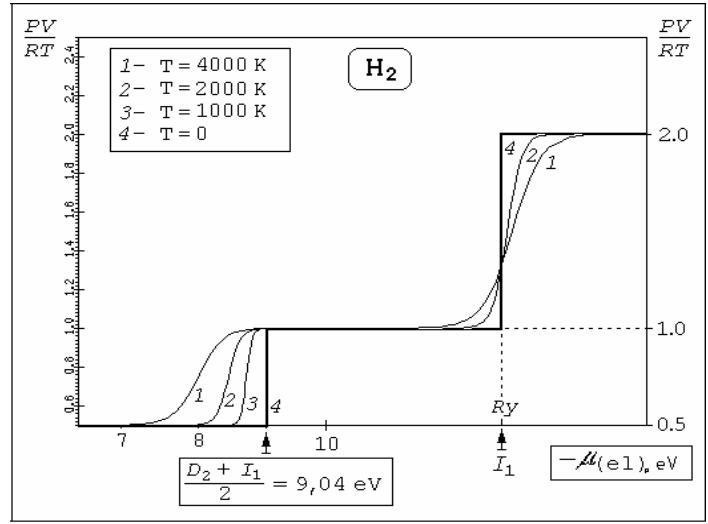

figure 3. Thermal EOS of hydrogen plasma in quasi-chemical limit (figure from [4]) Compressibility factor $PV/RT \equiv P/(n_{Li}kT)$ as a function of (negative) value of electron chemical potential. <u>Notations</u>: *1,2,3,4* – isotherms $T$ = 4000, 2000, 1000, 0 K correspondingly. $I_1$ = Ry – hydrogen ionization potential. $(D_2 + I)/2$ – position of "dissociation step" at electron chemical potential scale. Isotherms *1-3* are calculated via code SAHA [11] with neglecting of equilibrium radiation contribution.

This limiting structure appears within a fixed (negative) range of $\mu_{el}$ ($\mu_{el}^{**} \geq \mu_{el} \geq \mu_{el}^{*}$). It is bounded below by value of major ionization potential ($\mu_{el}^{*} = -I_Z = -Z^2 Ry$) and above by the value depending on ionization potential and sublimation energy of substance $\{\mu_{el}^{**} = -(\Delta_o H^S + I_1)/2\}$. Binding energies of all possible bound complexes (atomic, molecular, ionic and clustered) in its *ground state* are the *only* quantities that manifest itself in meaningful details of this limiting picture as location and value of every step. The energy of *macroscopic* binding – the heat of condensation at $T = 0$ – supplement this collection. At the same time there are *no* such *steps* for *exited states* of such bounded complexes (ions, atoms, molecules and clusters). Altogether, all energies mentioned above form "*intrinsic energy scale*" [3][10] for any substance.

In the zero-temperature limit all thermodynamic differential parameters (heat capacity, compressibility, *etc.*) obtain their remarkable δ-like structures ("thermodynamic spectrum" [3][10]). Both kinds of such "spectrum" became apparent: i.e. "emission-like spectrum" for heat capacity (fig. 3) and "absorption-like spectrum" for the isentropic coefficient - $(\partial \ln P / \partial \ln V)_S$ (fig. 4). It should be stressed again that all "lines" of these "thermodynamic spectrum" are centralized just at the elements of the "*intrinsic energy scale*" – binding energies of ground states for all bound complexes in the system.

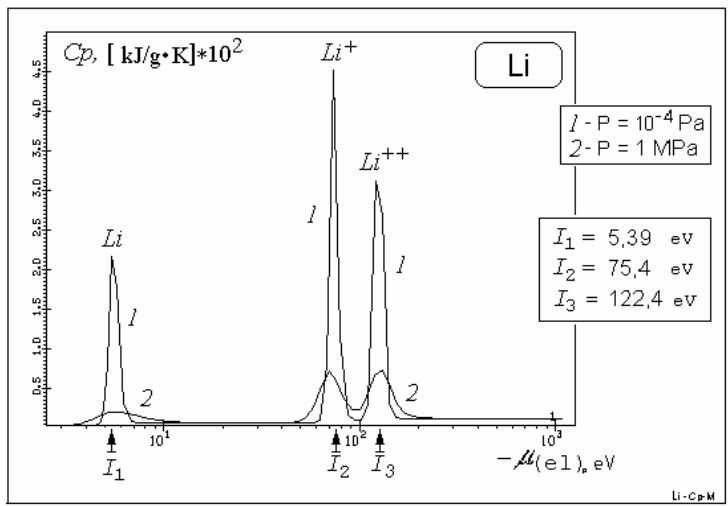

Figure 3. Limiting structure for differential thermodynamic quantities ("thermodynamic spectrum") in quasi-chemical limit $T \to 0$ (figure from [4][6]. Isobaric heat capacity of lithium plasma as a function of (negative) value of electron chemical potential. <u>Notations</u>: *1, 2* – isobars ($10^{-4}$ Pa and 1 MPa); arrows – elements of "*intrinsic energy scale*" for lithium: $I_1, I_2, I_3$ – lithium ionization potentials. (Isobars *2, 3* are calculated via code SAHA-IV [9] with neglecting of equilibrium radiation contribution)

The limiting EOS stepped structure ("ionization stairs") of gaseous zero-Kelvin isotherm is generic prototype of well-known "shell oscillations" in EOS of gaseous plasmas at low, but finite temperatures and non-idealities [2]. At the same time this limiting form of plasma thermodynamics could be used as a natural basis for rigorous deduction of well-known quasi-chemical approach ("chemical picture") in frames of asymptotic expansion around this reference system. The point is that this expansion must be provided on temperature at fixed chemical potential, in contrast to the standard procedure of expansion on density at constant temperature [12][1].

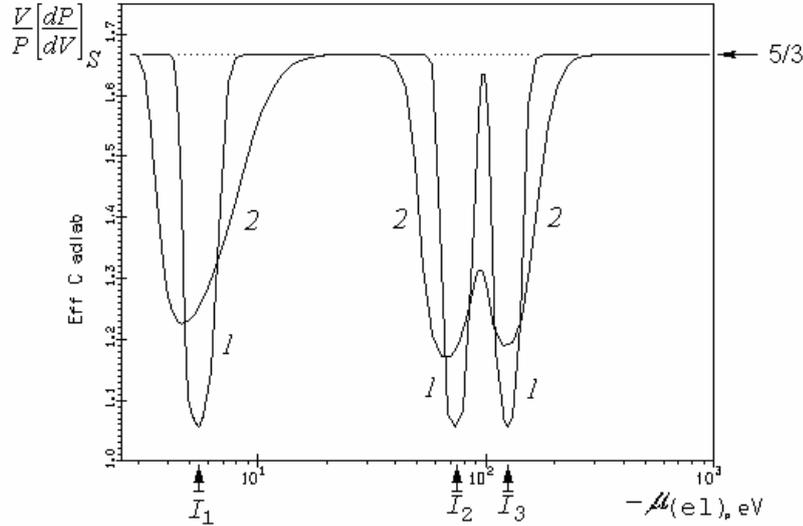

Figure 4. Limiting structure for differential thermodynamic quantities ("thermodynamic spectrum") in quasi-chemical limit $T \to 0$ (figure from [4][6][10]). Isentropic coefficient of lithium plasma $(\partial \ln P/\partial \ln V)_S$ as a function of (negative) value of electron chemical potential. <u>Notations</u>: – as at figure 3. Ideal-gas value $(\partial \ln P/\partial \ln V)_S = 5/3$ is noted.

The *gaseous branch* of zero-Kelvin isotherm $U_0^{gas}(\mu)$ could be naturally conjugated with associated *condensed branch* $U_0^{crystal}(\mu)$. Due to the choice of chemical potential as a ruling parameter this combination creates complete and *totally meaningful* non-standard "cold curve" for any substance $\{U_0(\mu)$ instead of $U_0(\rho)\}$. The point is the appearance of stable thermodynamic gaseous branch for this "cold curve", which reflects schematically *all reactions* (ionization, dissociation etc.) and phase transitions which are realized at the system. Besides, the stable part of new combined "cold curve" it could be supplemented with additional *metastable* branches, corresponding to overcooled vapour from gaseous part, and extended crystal from condensed part (figure 5) [13]. Another advantage of new representation for cold curve is natural identity of all transformations mentioned above (ionization, dissociation and phase transitions). It approve widely used interpretation of finite temperature ionization and dissociation as a "smoothed" phase transitions [10].

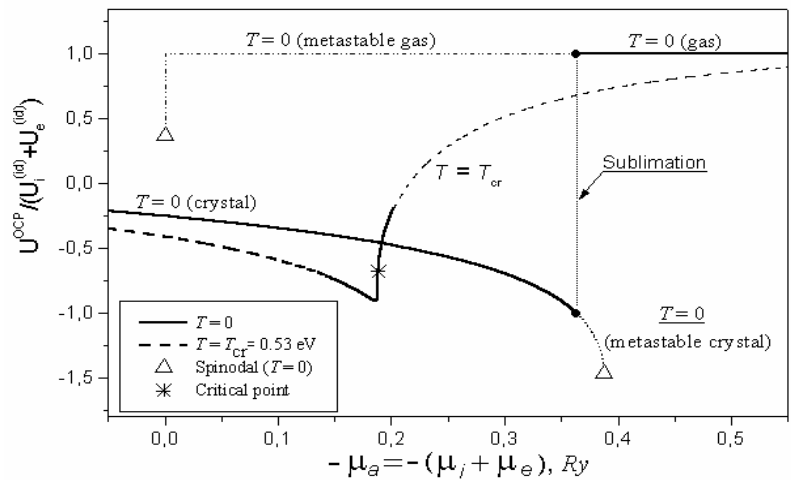

Figure 5. Modified "cold curve" (isotherm $T = 0$) and critical isotherm ($kT = kT_{cr} \approx 0.53$ eV) in modified one-component plasma model on *uniformly-compressible* compensating background $\{OCP(\sim)\}$ [14]. Dimensionless internal energy, $U^{OCP}/(U_i^{(id)} + U_e^{(id)})$, as function of "atomic" chemical potential (electroneutral combination of chemical potentials for ion and electrons (background)). Two metastable branches are exposed – *over-compressed* vapour and *extended* solid – completed by spinodal points (fig. from [3][10]).

Remarkable limiting structure of thermodynamics for real substances, which is under discussion, could manifest itself also in simplified classical models. Similar "ionization stairs", "thermodynamic spectrum" and modified "cold curve" was predicted [3][13][10] for modified one-component plasma on uniformly-compressible compensating background OCP(~) [14] [15], two-component classical ionic model with Glauberman's [16] potential $\{V_{ij}(r) \equiv Z_i Z_j e^2[1 - \exp(-r/\sigma)]/r\}$ [4], and for classical charged hard- and soft-spheres model. In the first case there is no electron-ionic associations in OCP(~) (on definition). The only transformations permitted in the model are 1$^{st}$-order phase transitions between solid, liquid and gas-like states. Non-standard cold curve of OCP(~) with sublimation jump and metatstable portions are shown at fig. 5.

All present statements about remarkable limiting structure of thermodynamic functions in zero-temperature limit for single substances are valid also in application to the chemical compounds. In this case one-dimensional structures: "ionization stairs", modified "cold curve" and "thermodynamic spectrums" turn into more complicated two-dimensional figures composed from discontinuities (steps) and ideal-gas planes. Features and properties of such limiting structures are non-investigated at the moment.

New representation for cold curve (isotherm $T=0$), which is introduced in present paper, has advantage for solution of theoretical problem of correct deducing of quasi-chemical representation (so-called "chemical picture" – ensemble of "free" simple and complex particles, atoms, molecules, ions and electrons with weak effective interaction) from rigorous physical representation (ensemble of nuclei and electrons with strong Coulomb interaction). Both "ionization stairs" in thermal and caloric EOS are natural zero-order terms in systematic asymptotic expansion for thermodynamic functions in the limit $T \to 0$ by the small parameter $\lambda \sim \exp\{-const/T\}$ at constant value of chemical potential [3][4][10]. It should be noted that well-known presently accepted traditional theoretical approach uses asymptotic expansion in terms of *activities* at *constant temperature* (for example [12]). Rigorous asymptotic expansion by functions of *temperature* in the limit $T \to 0$ (SAHA-limit) is developed for hydrogen in papers [7] [8] et al. for the region of atomic chemical potential corresponding to the case of electron-ion-atomic plasma. It should be stressed that desirable approach should develop asymptotic expansion in the limit $T \to 0$ simultaneously for *all values* of chemical potential within mentioned above "energy scale" from the state of full ionization ($\mu_{el} \sim -I_Z$) up to the saturation point and including the regions of *all stages* of ionization and atoms and molecules formation.

**Acknowledgments**

The work was supported by Grants: ISTC 3755 and by RAS Scientific Programs "Research of matter under extreme conditions" and "Physics of high pressure and interiors of planets"

**References**

[1] Iosilevski I. *High Temperature* **19** 494 (1981).
[2] Iosilevski I. and Gryaznov V. *High Temperature* **19** 799 (1981).
[3] Iosilevskiy I. *Plasma Thermodynamics in Zero-Temperature Limit* / Int. Conference "Physics of Non-Ideal Plasmas", Greifswald, Germany, 2000.
[4] Iosilevskiy I. / in *Encyclopedia of low-temperature plasma physics,* Supplement III-1 /eds. A. Starostin and I. Iosilevskiy, FIZMATLIT: Moscow, 2004, PP 349-428 *(in Russian)*
[5] Iosilevskiy I. / *Thermodynamic properties of low-temperature plasmas,* in: "*Encyclopedia of low--temperature plasma physics*" / ed. V. Fortov, Moscow: NAUKA, 2000, V.1, PP.284-288 (in Russian)
[6] Iosilevskiy I. *Thermodynamics of gaseous plasmas at the zero-temperature limit* / in "Physics of matter under extreme conditions" /ed. V. Fortov, IPCP RAS: Moscow, 2002, P.157 (in Russian).
[7] Brydges C. and Martin P. *J. Stat. Phys.* **96** 1163 (1999).
[8] Alastuey A., Ballenegger V. *J. Phys. A: Math. & Theor.* **42** (2009) (in press).
[9] Gryaznov V., Iosilevskiy I., Fortov V. *Thermodynamic properties of shock-compressed plasmas* / in: "*Encyclopedia of Low-Temperature Plasma Physics",* Supplement III-1 /eds. A. Starostin and I. Iosilevskiy, FIZMATLIT: Moscow, 2004, PP. 111-139 (*in Russian*)
[10] Iosilevskiy I., Krasnikov Yu., Son E., Fortov V. *Thermodynamic properties of non-ideal plasmas* / in "*Thermodynamics and Transport in Non-Ideal Plasmas*", MIPT Publishing: Moscow, 2000, PP.120-138 (in Russian); FIZMATLIT: Moscow, 2009, (in press).
[11] Gryaznov V., Iosilevskiy I., Krasnikov Yu., Kuznetsova N., Kucherenko V., Lappo G., Lomakin B., Pavlov G., Son E., Fortov V. // *Thermophysics of Gas Core Nuclear Engine* / Ed. by V.M. Ievlev, ATOMIZDAT: Moscow, 1980 (In Russian)
[12] Ebeling W., Kraeft W.D., Kremp D. *Theory of Bound States and Ionization Equilibrium in Plasmas and Solids /* Akademic–Verlag: Berlin, 1976).
[13] Iosilevskiy I. *Generalized "cold curve" and thermodynamics of a substance at the zero-temperature limit* // in "Physics of matter under extreme conditions" /Ed. V. Fortov, IPCP Chernogolovka: Moscow (2001) p.116 (in Russian).
[14] Iosilevskiy I. *High Temperature* **23** 807 (1985) arXiv:0901.3535
[15] Iosilevski I. and Chigvintsev A. /in "*Physics of Non-Ideal Plasmas"*// eds. W. Ebeling and A. Forster, Teubner: Stuttgart-Leipzig, 1992, PP. 87-94, arXiv:physics/0612113
[16] Glauberman A.E., *Doklady of Soviet Academy of Science,* **78**, 883 (1951)